\documentclass[onecolumn]{ceurart}
\usepackage[dvipsnames,svgnames]{xcolor} % text color
\usepackage[normalem]{ulem} % wavy underlines

\makeatletter
\font\uwavefont=lasyb10 scaled 700
\def\spelling{\bgroup\markoverwith{\lower3.5\p@\hbox{\uwavefont\textcolor{Red}{\char58}}}\ULon}
\def\grammar{\bgroup\markoverwith{\lower3.5\p@\hbox{\uwavefont\textcolor{LimeGreen}{\char58}}}\ULon}
\def\phrasing{\bgroup\markoverwith{\lower3.5\p@\hbox{\uwavefont\textcolor{RoyalBlue}{\char58}}}\ULon}

\newcommand\remove{\bgroup\markoverwith{\textcolor{red}{\rule[0.5ex]{2pt}{0.4pt}}}\ULon}
\newcommand\insertion{\bgroup\markoverwith{\textcolor{Green}{\rule[-0.5ex]{2pt}{0.6pt}}}\ULon}
\makeatother

\usepackage[inline]{enumitem}
\newlist{inlinelist}{enumerate*}{1}
\setlist[inlinelist,1]{label=\textit{\roman*)}}

\usepackage{minted}
\setminted{breaklines=true}

\begin{document}

\copyrightyear{2025}
\copyrightclause{Copyright for this paper by its authors. Use permitted under Creative Commons License Attribution 4.0 International (CC BY 4.0).}

% TODO:
\conference{3nd Solid Symposium, 24--25 April 2025, Leiden}

%% TODO:
\title{From Access Control to Usage Control with User-Managed Access}

%% TODO:
\author[1]{Wout Slabbinck}[%
orcid=0000-0002-3287-7312,
email=wout.slabbinck@ugent.be,
url=https://woutslabbinck.com/
]
\cormark[1]
\address[1]{IDLab, Department of Electronics and Information Systems,
Ghent University - imec, Belgium }
\author[1]{Wouter Termont}[
orcid=0000-0002-2968-1394,
email=wouter.termont@ugent.be
]
\author[1]{Ruben Dedecker}[
orcid=0000-0002-3257-3394,
email=ruben.dedecker@ugent.be,
url=https://rubendedecker.be/
]
\author[1]{Beatriz Esteves}[
orcid=0000-0003-0259-7560,
email=beatriz.esteves@ugent.be,
url=https://w3id.org/people/besteves
]
% %% Footnotes
\cortext[1]{Corresponding author.}

%%
%% The abstract is a short summary of the work to be presented in the article.
%%
%% The abstract is a short summary of the work to be presented in the article.
\begin{abstract}
% Context:      Why the need is so pressing or important
%Solid has vision for control over data. For that we need usage control.
Recent data protection and data governance regulations intensify the demand for interoperable, decentralized data ecosystems
that can support not only access control but also legally-aligned governance over data use.
% Need:         Why something needed to be done at all
%To achieve that, Solid envisioned to use UMA (https://solidproject.org/TR/oidc-primer\#request-flow)
Existing Web-based data storage platforms increasingly struggle to meet these regulatory and practical requirements,
as their authorization mechanisms rely on tightly coupled, document-centric access control models
that lack expressiveness for legal constraints and fail to separate data management from authorization concerns.
In parallel, widely adopted authorization standards remain poorly aligned with decentralized, semantically rich usage-control scenarios.
% Task:         What was undertaken to address the need
%We have created a UMA server that works with the Community Solid Server
To bridge this gap, this work introduces an architecture that replaces Solid's native access control mechanisms with a User-Managed Access~(UMA) authorization flow, enabling the enforcement of usage control policies expressed with the W3C Open Digital Rights Language~(ODRL) standard.
% Object:       What the present document does or covers
%the UMA AS allows for interactive claims gathering (allowing VC and DID above WebID),
%allows for a custom policy engines (such as an ODRL one)
In this context, this article details the conceptual background motivating this approach, presents the proposed UMA-based architecture, and describes a prototype implementation that integrates an ODRL-enabled Authorization Server with a Solid-compatible Resource Server.
% Findings:     What the work done yielded or revealed
The prototype demonstrates that decoupling authorization from storage enables more flexible, interoperable, and legally expressive control over data use, while remaining compatible with existing Solid infrastructure. It also highlights practical design choices required to evaluate ODRL policies in the absence of a fully standardized evaluation semantics.
% Conclusion:   What the findings mean for the audience
Moreover, this work shows how usage control can be operationalized using existing Web standards, offering a concrete path beyond permission-based access control toward policy-aware, legally informed data governance.
% Perspectives: What the future holds, beyond this work
Future research will focus on policy management interfaces for resource owners, richer claim verification mechanisms, and techniques for communicating and enforcing obligations over time.
\end{abstract}

% \begin{keywords}
%   \LaTeX{} class \sep
%   paper template \sep
%   paper formatting \sep
%   CEUR-WS
% \end{keywords}

\maketitle

\section{Introduction}\label{sec:intro}

With legislative efforts such as the European Union's General Data Protection Regulation (GDPR)~\cite{gdpr} and Data Act~\cite{data_act} regulating rights and obligations around data portability, human self-determination, and data reuse, there is a growing effort by businesses and other data-oriented organizations to enable data interoperability and governance.
In this context, Semantic Web technologies emerge as strong enablers as they offer open, machine-interpretable standards that support decentralization and cross-platform integration.
In the past two decades, several projects have matured that attempt to leverage Semantic Web technologies to achieve broad interoperability between systems for storage, access and (re-)\,use of resources in a global, decentralized ecosystem.
Most notable among them are Fedora\footnote{The \emph{Flexible Extensible Digital Object Repository Architecture} (Fedora, https://fedora.info)~\cite{fedora}.} and Solid\footnote{Solid (pseudo-acronym, originally \emph{Social Linked Data}; https://solidproject.org/)~\cite{sambra_solid_2016}.}, two projects that have been taken up as input in the World Wide Web Consortium's \emph{Linked Web Storage} Working Group (W3C LWS WG), to consolidate these efforts in a \emph{W3C Recommendation} web standard~\cite{lws}.

Since they originate from the 2000s, before the dawn of modern access control,
%\be{what are we considering as the start of modern access control here and how does it differ from `traditional' access control?}
%\ws{I think it is related to the fact that the web originally had no proper access control and this was before the oauth initiatives.
%Maybe also the fact that access control often is focused on central systems rather than decentralized.
%Though I find papers from 2005 about ABAC (which is distributed by design) on the web (though not semweb)\cite{yuan_AttributedBasedAccess_2005}. Also ABAC existing already in 1997 \cite{servos_CurrentResearchOpen_2017}. There is always the survey of Sabrina that we could use to give some context why these two policy frameworks were used \cite{kirraneAccessControlResource2017}}
both projects ended up using custom access control frameworks: Web Access Control (WAC)~\cite{noauthor_web_nodate} and ---in the case of Solid--- Access Control Policies (ACP)~\cite{noauthor_acp_nodate}, two specifications that prescribe a language for writing lists of policies,
and an algorithm for evaluating access requests in accordance with those policies. Almost parallel to these developments, however, the limited flexibility of the HTTP \texttt{Authorization} header ---the main access control mechanism for HTTP, for which it was initially a non-goal~\cite{berners_lee_information}--- and the increased vendor lock-in caused by proprietary `solutions', provided the spark for OAuth~\cite{rfc6749}, today's \emph{de facto} standard for access control on the Web.

This parallel evolution of more elaborate access control frameworks, like OAuth, with semantically interoperable decentralized storage systems, like Fedora and Solid, has resulted in a situation where the former are typically restricted to centralized, uninteroperable use-cases, while the latter are unable to handle more advanced authorization scenarios.
An important class of such scenarios involves legal constraints and, more broadly,
the consideration of prohibitions and obligations to be as equally important statements as permissions, in the definition of what can be done with a certain resource.
The capacity to semantically express such ``rules'' broadens a limited view on access control ---a one-off ``who can access what''--- to the more holistic concept of \emph{usage control}: the rights and duties of each party throughout the lifetime of the data~\cite{slabbinck_enforcing_usage_control_2024}. While there exist frameworks for modeling such requirements, e.g., W3C's Open Digital Rights Language (ODRL)~\cite{iannella_odrl_2018} and Data Privacy Vocabulary (DPV)~\cite{pandit_dpv}, these are not easily integrated into WAC or ACP~\cite{esteves_odrl_profile, florea_is_2023}.

In this paper, we propose one way to address this discrepancy: we enable ODRL policies to protect the Solid storage API as a replacement for WAC/ACP.
We chose ODRL since it is an interoperable W3C standard for expressing usage-control policies, which can be extended through profiles, for example by combining DPV and ODRL to create legally-aligned policies.
Since in Solid WAC and ACP are tightly coupled to the storage layer, we adopt an OAuth~2.0-based authorization-API, using the User-Managed Access (UMA) extension, to separate authorization from storage interactions.
After elaborating in Section \ref{sec:background} on the background, terminology of the different involved frameworks and on the selection of UMA, we give an overview of the architecture we propose in Section \ref{sec:architecture}, and in Section \ref{sec:implementation} present a prototype implementation of it. 
In Section \ref{sec:discussion}, we then compare our framework with WAC and ACP, discuss to what extent it overcomes their limitations and identify areas where future work can refine and extend the implementation.
Finally, we conclude with several suggestions for future research in Section \ref{sec:conclusion}.
\section{Background}\label{sec:background}

An elaborate overview of the shortcomings of WAC and ACP can already be found in~\cite{solidlab:uma}. The majority of the issues follows from a lack of separation between orthogonal concerns: management of data, and management of control.

An obiquitous characteristic of modern access control mechanisms, like OAuth, is a strict separation between Resource Servers (RS) ---the ``data plane''--- and Authorization Servers (AS) ---the ``control plane''. Since WAC and ACP were drafted before this design pattern widespread, they let one party (the RS) incorporate both concerns. While architecturally simpler, since the protection domain can be tailored to a single storage API, this is inflexible from the perspectives of management and auditing. More importantly, however, it has concrete impact on privacy and security. For example, all information (e.g., user identity) is unnecessarily revealed to a single party; and the responsibility of technical evolution (e.g., the upgrade path of authentication methods) is scattered over each individual server.

One major consequence is that WAC and ACP are tightly coupled to the Linked Data Platform~(LDP)~\cite{speicherLinkedDataPlatform2015}, the storage API provided by both Solid and Fedora. LDP handles data as a hierarchy of documents, which assumes a symmetry between how data is written and how it is subsequently read\footnote{A more in-depth analysis of this read--write symmetry can be found in~\cite{dedecker_whats_in_a_pod_2022}.}. WAC and ACP inherit these assumptions: they are restricted to symmetric, per-document policies, and cannot protect data with other structures. This dependency requires applications to know about the existence of data before knowing whether they can access that data, while ideally the opposite would be true.

Similarly, since the WAC and ACP policy languages are tightly integrated with their authorization algorithms, and at the same time exposed as their interface to external applications, it is difficult to update or change any of those aspects. For example, Slabbinck et al.~\cite{slabbinck_enforcing_usage_control_2024} demonstrate the complications of integrating WAC or ACP with ODRL, in an attempt to model legal and user requirements.

% Each of these papers emphasizes the importance of usage control over mere access control:
% \footnote{While access control is merely concerned with which parties can access what resources, usage control includes  the conditions and obligations associated with this authorization.}

% 23: openid_connect
% 24: w3c_odrl
% 25: slabbinck_enforcing_usage_control_2024
% 26: esteves_odrl_profile
% 27: w3c_dpv
% 28: pandit_dpv
% 29: dedecker_whats_in_a_pod_2022
% 31: berners_lee_information_management
% 32: ietf_oauth21

\subsection{OAuth 2.x}\label{sec:oauth}

The clean separated architecture of OAuth~\cite{rfc6749}---in particular version 2.x--- largely avoids the problems listed above; but this does not mean that its design is without faults. Two main issues block the way towards a decentralized and interoperable ecosystem of data storage and (re-)\,use.

First, despite its separation of data and access control, OAuth's design is aimed at \emph{static}, \emph{insulated} protection domains: one AS manages access to one or more preestablished RSs. Not unlike WAC and ACP, this precludes scenarios in which one wants certain policies to apply to multiple specific resources, spread over a variety of different RSs ---a key requirement for decentralized, user-managed systems.

Second, OAuth relies on purely \emph{synchronous} flows, in which applications and their users either have access or not. At least in core OAuth 2.x, or any of its officially endorsed extensions, it is not possible ---\emph{in-band}--- to \emph{request} the Resource Owner (RO) to change their policies. This not only drastically limits its immediate usefulness in many applications; it also poses a serious challenge to automation.

\subsection{User-Managed Access}\label{sec:uma}

The Kantara Initiative's User-Managed Access (UMA) \cite{maler_user-managed_2018} extension to OAuth 2.0 addresses these limitations. It enables asynchronous access negotiation between the RO and third parties, with support for interactively requesting them to present a variety of claims.
Moreover, it allows for dynamic federation of control over multiple protection domains.

These insights are not new: UMA's inspiration has been taken up by the Grant Negotiation and Authorization Protocol~(GNAP)~\cite{gnap}, and an attempt has even been made to introduce it in the Solid specification: since December 2021, Solid-OIDC~---~the project's only normative authentication protocol~---~recommends that the Solid-OIDC identity claim is exchanged for an OAuth access token at an UMA server~\cite{noauthor_solid-oidc_2022}; no other possible flow is suggested. Yet somehow, all Solid servers known to date still expect the identity token to be sent to themselves, for verification according to WAC or ACP. The confusion is understandable, since both specifications are normative parts of the Solid protocol~\cite{capadisli_SolidProtocol_2024}. 

Note that UMA, as a point of orthogonality, does not define a policy language, nor a decision algorithm. The expressiveness of the policy rules, in comparison with WAC and ACP, will thus depend on additional choices to fill these seats. An obvious choice would be the Solid Application Interoperability (SAI) specification~\cite{w3c_sai}, a set of data formats that overlay the hierarchical document-centric interface of LDP, to provide a more flexible management of \emph{types} of data and access to them. However, being merely a technical interoperability layer on top of WAC or ACP, the SAI specification still lacks the semantics to express legal constraints~\cite{esteves_semantic_2024}.

The state of the art framework to model such constraints is the ODRL W3C Recommendation~\cite{iannella_odrl_2018}. As an open foundation for modeling usage control rules over the full lifecycle of data use, it can handle a wide variety of use cases. In particular, when combined with DPV~\cite{pandit_dpv}, it can be used to express the legal grounds and purposes necessary to engage in modern-day data exchange.

% feedback from peer review:
% - comparative table of different approaches would be nice
% - terms should be defined at the start and used consistently
%   => most architectural ones are already in intro/background

\section{Architecture}\label{sec:architecture}

% TODO: focus more on actual UMA implementation}
% TODO: Remove the data minimization, that we have not implemented yet}
% Elaboration on UMA access flow before explaining the architecture
The UMA grant flow~\cite{maler_user-managed_2018} begins with the \textbf{Resource Owner (RO)},
who defines access policies for resources hosted on the \textbf{Resource Server (RS)} and managed by the \textbf{Authorization Server~(AS)}.
When a \textbf{Requesting Party (RP)}, via a \textbf{Client}, attempts an operation on a protected resource, the RS denies the initial request due to the absence of a Requesting Party Token (RPT).
Instead, it provides the Client a permission ticket and the AS endpoint URI\footnote{Following the Federated Authorization for User-Managed Access (UMA) 2.0 specification~\cite{maler_fed-UMA_2018}, the RS obtains this ticket from the AS before responding to the Client.}.

The Client then interacts with the AS by providing the ticket and the RP's identity claims. 
The AS performs an authorization assessment by evaluating the RO's policies against these claims; if successful, it issues a Requesting Party Token (RPT). 
Finally, the Client re-attempts the operation on the protected resource, this time presenting the obtained RPT. 
% Finally, the Client retries again to perform the action on the protected resource, but now provides the RPT.
The RS validates this token and, having confirmed the authorization, allows the request to proceed.

\begin{figure*}[ht]
    \centering
    \includegraphics[width=\textwidth]{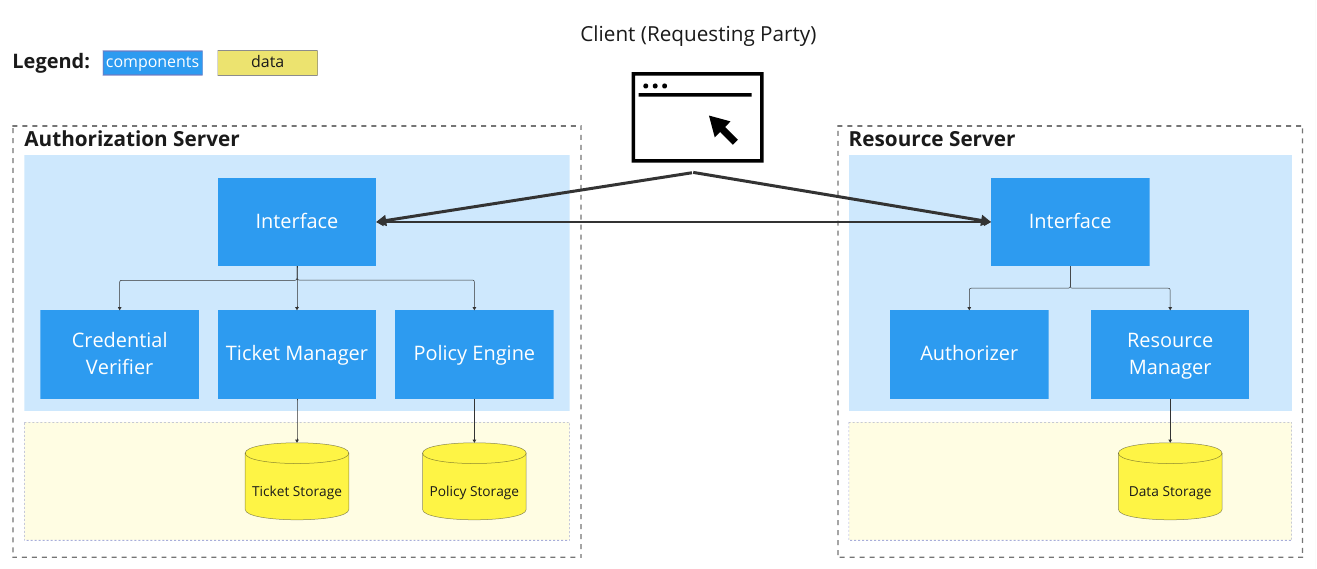}
    \caption{Two-tier architectural overview depicting the UMA Authorization Server and Resource Server.}
    \label{fig:architecture}
\end{figure*}

Building on this flow, our architecture, illustrated by Figure \ref{fig:architecture}, focuses on adhering to the UMA specification and the ability to enforce usage control policies.
As such, we designed an architecture with a focus on the Authorization Server and the Resource Server.
% The \textbf{Authorization Server (AS)} is responsible for 
% \begin{inlinelist}
%     \item checking data access requests, 
%     \item verifying user credentials, 
%     \item evaluating usage control policies and 
%     \item issues Requested Party Token (RPT) that grant access to resources.
% \end{inlinelist}
The \textbf{Authorization Server} consists of four main components, respectively an Interface, a Ticket Manager, a Policy Engine, and a Credential Verifier, and two storage components.
The \textbf{Interface} provides the UMA AS interface, handling requests with a ticket and claims. 
It orchestrates with the other components in three steps to, when all successful, return an RPT. 
First, it queries the \textbf{Ticket Manager} using the ticket to obtain the requested permissions from the \textbf{Ticket Storage}. 
Next, it provides the claims to the \textbf{Credential Verifier} for validation. 
Finally, it transmits the claims and requested permission to the \textbf{Policy Engine}, which then fetches relevant policies from the \textbf{Policy Storage} to make an access decision.

The \textbf{Resource Server} consists of three components.
The \textbf{Interface} provides the UMA RS interface, responding to client requests by either redirecting them with a ticket when no RPT is present or allowing the operation on the protected resources when a valid RPT is provided.
It delegates validation of the RPT against the requested operation and requesting tickets to the \textbf{Authorizer} component.
To allow the operations, the Interface uses the \textbf{Resource Manager} to mediate access to the \textbf{Data Storage}. The Resource Manager ensures that all HTTP operations are performed in a consistent, fault-tolerant manner.

\section{Implementation}\label{sec:implementation}

Adhering to the aforementioned architecture, depicted in the Section~\ref{sec:architecture},
we implemented an open-source UMA server prototype governing usage control to Solid servers,
which is available at \url{https://github.com/SolidLabResearch/user-managed-access/}
under the MIT license.
The remainder of this section expands further implementation details of the two mentioned servers.

\subsection*{Resource Server}
Built upon the modular Community Solid Server (CSS)~\cite{van_herwegen_community_2024}, our implementation specializes the server's internal authorization flow. 
By replacing the default access control layer with a custom RPT-driven authorizer, we adapted the CSS to be a fully compliant UMA Resource Server while maintaining its native Linked Data Platform (LDP) interface.

The \textbf{Authorizer} manages access through three phases: permission calculation, token validation, and contingent ticket issuance.
It first translates the LDP operation into \textbf{required permissions} (Resource URI and Access Right tuples). 
It then attempts to validate that the provided RPT is authentic and contains all required permissions.
If validation fails or the RPT is missing, the Authorizer requests a permission ticket from the AS. 
Successful authorization proceeds to the LDP operation, while failure returns a \texttt{401} status with the ticket and AS location in the \texttt{WWW-Authenticate} header.
As such, the authorization decision is completely delegated to the appropriate Authorization Server.

\subsection*{Authorization Server}
An UMA Authorization Server is responsible for guarding access to protected resources of an UMA Resource Server on behalf of a Resource Owner.
We elaborate in this section how we implemented an UMA compliant Authorization Server that enforces usage control using ODRL policies.
Thereby we focus on the happy flow\footnote{We focus on the happy flow to illustrate the interaction of the main components and the integration of policy evaluation; alternative branches are implemented but omitted for brevity.}, specifically the process of a Requesting Party acquiring an RPT as defined in Sections 3.3.1, 3.3.4, and 3.3.5 of the UMA specification~\cite{maler_user-managed_2018}.

% \todo{ does this require some introduction? Note that we currently do not focus on RS requesting tickets and resource registration -> mention that we just follow straight UMA stuff for that and we focus on UMA 3.3.4 Policy assessment and result determinitan}

When handling a request for an access token, four core steps are executed sequentially:
\begin{inlinelist}
    \item parsing the RPT request,
    \item figuring out the requested permissions,
    \item extracting and verifying the claims, and
    \item assessing authorization
\end{inlinelist}.
These steps are detailed in the next subsections.

%% NOTE: current problem, this only solves step 3.3.2
\paragraph*{Request Parsing}
\begin{sloppypar} % permissions in the second sentence was not breaking nicely
    This subsection concerns the parsing of the incoming HTTP request sent to the AS by a RP Client.
    The parser ensures the \texttt{grant\_type}\footnote{The UMA specification mandates the presence of the fixed grant type \texttt{urn:ietf:params:oauth:grant-type:uma-ticket}.} alongside either a \texttt{ticket} or a \texttt{permissions} list are present in the HTTP body.
    Although UMA 2.0 requires a \texttt{ticket}, we additionally support the \texttt{permissions} field as introduced by \cite{solidlab:uma},
     where direct mode is specified to let the RP avoid the RS round-trip when the AS is known.
    Furthermore, the parser extracts optional fields such as \texttt{claim\_token} and \texttt{claim\_token\_format}, which support client authentication and allow different types of authentication claims to be handled.
\end{sloppypar}

%% Only when the ticket exists
\paragraph*{Determining Requested Permissions}
The ticket manager resolves the token access request into a concrete set of permissions for the requesting party. 
It first verifies the existence of the provided ticket; if valid, the associated permissions are retrieved. 
If no ticket exists, the requested permissions were supplied through the direct mode.

\paragraph*{Verifying the claims}
We utilize the modularity of Components.js~\cite{taelman_swj_2023} to implement a typed verification layer that supports diverse credential formats 
based on the \texttt{claim\_token\_format}. 
Currently, we provide a specialized verifier for OIDC 2.0 Tokens, to ensure strict alignment with the Solid-OIDC protocol. 
By validating these tokens, the server transforms raw credentials into a set of trusted claims anchored by a verified WebID. 
This resulting claim set serves as the authoritative, verified context for the subsequent policy evaluation.

\paragraph*{Authorization assessment} The Authorization Server's \textbf{policy engine} is responsible for the assessment of the policies.
Since the UMA specification does not impose any requirements on the policy language and we want to be able to enforce usage control, ODRL is chosen as the policy language. As such, the evaluation of policies entails
\begin{inlinelist}
    \item transforming verified claims and requested permissions into formal Evaluation Requests,
    \item evaluating these requests against the ODRL policies, and
    \item transforming the outputs of the evaluation back to permissions (which the Resource Server understands).
\end{inlinelist}
For each incoming RPT request, the engine derives an Evaluation Request that links the RP's claims to the permissions being requested. 
%\ws{actually, if there are multiple actions requested for a resource or multiple resources. Multiple will be generated since the evaluator currently only handles one (subject, target, resource) evaluation request at the same time. Perhaps in a future iteration, we could incorporate semantics on what should happen when multiple are supplied}
It also generates the corresponding State of the World, which captures contextual conditions such as the current timestamp. Both structures follow the models introduced in \cite{estevesCapturingRequestsContext2025}. 
These two inputs, together with the ODRL policies, are then supplied to the ODRL evaluator\footnote{As ODRL Evaluator, we employ the one introduced in \cite{wout_slabbinck_interoperable_2025}.}.
%\be{cite the evaluator and the inputs paper? link to our evaluator as a footnote?}\ws{I've added it. A side note is that currently, this is not done that way and probably wont be exact like how we introduced it there.}\be{I think it is fine, we say that we use the models and evaluator introduced in the publications and not the exact version of the publications, of course these things will always have new versions and the sentence does not contradict that.}

The ODRL Evaluator produces Compliance Reports\footnote{
We follow the Compliance Report model introduced in \cite{wout_slabbinck_interoperable_2025} to elaborate the result of an ODRL evaluation process. The model is openly available at \url{https://w3id.org/force/compliance-report}.}.
To decide whether access should be granted or not, the engine then interprets these Reports.
In this context, a Compliance Report may contain multiple Rule Reports, each representing the outcome of evaluating a single ODRL rule. 
A Rule Report is active only when all its premises hold. An active Permission Report supports granting the requested RPT, 
while an active Prohibition Report requires denying access. Inactive reports are ignored.
Since a Compliance Report may contain only inactive Rule Reports or may include conflicting active ones (both Permission and Prohibition), 
the engine applies two resolution strategies: \textbf{`default-deny'} and \textbf{`prohibition-overrides-permission'}.
`Default-deny' ensures that no access is granted when there are no active Rule Reports. 
The `prohibition-overrides-permission' strategy resolves conflicts by giving precedence to prohibitions.
As a result, a permission is granted only if at least one active Permission Report supports it and there are no active Prohibition Reports. 
Finally, based on these Reports, the authorized permissions are serialized into an RPT and returned to the Requesting Party Client.

\section{Discussion}\label{sec:discussion}
The implementation of the UMA specification for a Solid server separates the concerns of handling reading and writing resources from verifying whether agents are authorized to do so.

% impact separation of concerns: no hierarchies -> no collections
Since UMA does not specify how to exchange collection information, hierarchical or other set-based information, that is an artifact from the RS interface, is not taken into account during policy evaluation.
Compared to a Solid server using WAC or ACP-based authorization, this might prevent accidental data leakage when the Resource Owner failed to recall liberal access control policies in a parent container.

% impact use of ODRL -> we need formalisation
To ensure interoperability, we adopted the ODRL standard over adapting WAC or ACP to enforce usage control policies.
However, whereas the expressing of ODRL policies is standardized, the evaluation of these policies unfortunately is not.
This means there is no single algorithm for deterministic ODRL policy evaluation~\cite{bonattiFormalSemanticsOpen2025}.
As such, we build on top of the ODRL Evaluator~\cite{wout_slabbinck_interoperable_2025} and made some implementation choices that adheres to the formalization efforts of the ODRL Community Group\footnote{Besides maintaining ODRL's 2.2 Recommendation, the W3C's ODRL Community Group is defining the ODRL Formal Semantics specification at \url{https://w3c.github.io/odrl/formal-semantics/}, which details a deterministic algorithm for software implementing ODRL policy evaluation.} to incorporate a strategy to deal with conflicting rules.

% Monitoring enforcement needs more work
Furthermore, while ODRL allows you to describe obligations in policies, such as \textit{"Delete after 24 hours"}, the use of UMA does not directly allow you to monitor whether this is fulfilled or not.
This limitation is two-fold.
First, during the request for an access grant, the AS does not disclose the policies, only the opaque token itself.
Consequently, the client remains unaware of specific obligations.
Second, UMA lacks an endpoint for clients to provide proof of fulfillment. %also the client does not need to have an endpoint, so there is no way to send a notification to ask whether something was performed or not.
To bridge the gap, future work could entail a sticky policy mechanism where a specialized ODRL Policy is embedded within the access token. This ensures that the policy follows the data without compromising the actual policies of the RO.
Furthermore, a new endpoint could be introduced to gather proofs of fulfillment, mirroring the existing interactive claims-gathering mechanism.

% More generic claims
Expanding support for additional claim types would also further strengthen the Credential Verifier. 
Two promising directions are the verification of Verifiable Presentations~\cite{spornyVerifiableCredentialsData2025} and the inclusion of purpose claims. 
Verifiable Presentations would enable the system to validate claims issued by trusted authorities, which could be managed through an additional Verifiable Data Registry integrated into the Authorization Server. 
For example, the system could confirm a government-issued credential stating that an individual is a licensed medical doctor.
On the other hand, purpose claims would allow policy definitions to more closely reflect legal requirements such as those found in the GDPR~\cite{gdpr}.
Together, these capabilities would enable the enforcement of generic policies that permit health professionals to use data for research purposes.

% Need for LOAMA
Finally, a conventional Solid Server does not have the distinction between a Resource Server and an Authorization Server. 
Consequently, policies are managed as standard data resources; changing access rights requires directly modifying the auxiliary resources associated with a given data resource, such as the~\texttt{.acl} or~\texttt{.acp} files used by WAC and ACP, respectively.
In contrast, managing policies in UMA is done only via the Authorization Server.
Our work focuses strictly on the enforcement of ODRL policies within UMA.
Since the specification leaves the policy management interface up to the implementer, we left this out for future development.

\section{Conclusion \& Future work}\label{sec:conclusion}
In this paper, we introduced an implementation of a Solid-based UMA deployment, built on the Community Solid Server, that conforms with the Solid-OIDC resource access flow.
It decouples the combined data and control resource approach of Solid into respectively a Resource Server serving the LDP interface and an Authorization Server that handles both the authentication and authorization processes.
To support usage control enforcement, the Authorization Server employs ODRL policies to calculate access permissions.
Due to a lack of formalism for evaluation of ODRL policies, pragmatic choices were made to ensure consistent policy decisions through the use of, among others, ad-hoc strategies to handle policy conflicts.

Future work entails an implementation of a policy management interface for Resource Owners. 
Furthermore, research on agreement instantiation would make it possible to attach context-specific policy information directly to the data while keeping the Resource Owner's generic policies confidential. 
This would allow Requesting Parties to be informed of their concrete obligations without revealing sensitive policy information.
Finally, another research direction concerns how to represent different types of claims within ODRL in a way that preserves the separation of concerns, ensuring that the policy language supports required claims without shifting credential-verification responsibilities onto the ODRL Evaluators.

\begin{acknowledgments}
This research was funded by SolidLab Vlaanderen (Flemish Government, EWI and RRF project VV023/10)
and by the imec.icon project PACSOI (HBC.2023.0752), which was co-financed by imec and VLAIO
and brings together the following partners: FAQIR Foundation, FAQIR Institute, MoveUP, Byteflies,
AContrario, and Ghent University -- IDLab.
\end{acknowledgments}

%%
%% Define the bibliography file to be used
\bibliography{bibliography}

@article{sambra_solid_2016,
	title = {Solid: {A} {Platform} for {Decentralized} {Social} {Applications} {
	         Based} on {Linked} {Data}},
	language = {en},
	author = {Sambra, Andrei Vlad and Mansour, Essam and Hawke, Sandro and Zereba,
	          Maged and Greco, Nicola and Ghanem, Abdurrahman and Zagidulin, Dmitri
	          and Aboulnaga, Ashraf and Berners-Lee, Tim},
	year = {2016},
	pages = {16},
}

@misc{iannella_odrl_2018,
	title = {{ODRL Information Model 2.2 -- W3C Recommendation 15 February 2018}},
	url = {https://www.w3.org/TR/odrl-model/},
	author = {Iannella, Renato and Villata, Serena},
	year = {2018},
}

@inproceedings{wout_slabbinck_interoperable_2025,
	address = {Cham},
	title = {Interoperable {Interpretation} and {Evaluation} of {ODRL} {Policies}},
	isbn = {978-3-031-94578-6},
	doi = {10.1007/978-3-031-94578-6_11},
	booktitle = {The {Semantic} {Web}},
	publisher = {Springer Nature Switzerland},
	author = {Slabbinck, Wout and Rojas Meléndez, Julián and Esteves, Beatriz and Colpaert, Pieter and Verborgh, Ruben},
	editor = {Curry, Edward and Acosta, Maribel and Poveda-Villalón, Maria and van Erp, Marieke and Ojo, Adegboyega and Hose, Katja and Shimizu, Cogan and Lisena, Pasquale},
	year = {2025},
	pages = {192--209},
}

@misc{spornyVerifiableCredentialsData2025,
  title = {Verifiable {{Credentials Data Model}} v2.0},
  author = {Sporny, Manu and Longley, Dave and Chadwick, David and Herman, Ivan},
  year = 2025,
  publisher = {W3C Working Group},
  url = {https://www.w3.org/TR/vc-data-model-2.0/}
}

@article{van_herwegen_community_2024,
	title = {The {Community} {Solid} {Server}: {Supporting} research \&
	         development in an evolving ecosystem},
	copyright = {https://creativecommons.org/licenses/by/4.0/},
	issn = {22104968, 15700844},
	shorttitle = {The {Community} {Solid} {Server}},
	url = {
	       https://www.medra.org/servlet/aliasResolver?alias=iospress&doi=10.3233/SW-243726
	       },
	doi = {10.3233/SW-243726},
	language = {en},
	urldate = {2024-11-29},
	journal = {Semantic Web},
	author = {Van Herwegen, Joachim and Verborgh, Ruben},
	editor = {Hose, Katje},
	month = oct,
	year = {2024},
	pages = {1--15},
}

@misc{rfc6749,
	series = {Request for Comments},
	number = 6749,
	howpublished = {RFC 6749},
	publisher = {RFC Editor},
	doi = {10.17487/RFC6749},
	url = {https://www.rfc-editor.org/info/rfc6749},
	author = {Dick Hardt},
	title = {{The OAuth 2.0 Authorization Framework}},
	pagetotal = 76,
	year = 2012,
	month = oct,
}

@misc{noauthor_solid-oidc_2022,
	title = {Solid-{OIDC}},
	url = {https://solidproject.org/TR/oidc},
	urldate = {2025-03-13},
	year = {2022},
}

@misc{noauthor_web_nodate,
	title = {Web {Access} {Control} ({WAC})},
	url = {https://solidproject.org/TR/wac},
	urldate = {2022-01-21},
	year = {2022},
}

@misc{noauthor_acp_nodate,
	title = {{Access Control Protocol (ACP)}},
	url = {https://solidproject.org/TR/acp},
	urldate = {2022-05-18},
	year = {2022},
}

@misc{maler_user-managed_2018,
	title = {User-{Managed} {Access} ({UMA}) 2.0 {Grant} for {OAuth} 2.0 {
	         Authorization}},
	url = {
	       https://docs.kantarainitiative.org/uma/wg/rec-oauth-uma-grant-2.0.html#RFC6749
	       },
	language = {en},
	urldate = {2025-03-13},
	author = {Maler, E. and Machulak, M. and Richer, J.},
	month = jan,
	year = {2018},
}

@misc{maler_fed-UMA_2018,
	title = {{Federated Authorization for User-{Managed} {Access} ({UMA}) 2.0}},
	url = {
	       https://docs.kantarainitiative.org/uma/wg/rec-oauth-uma-federated-authz-2.0.html
	       },
	language = {en},
	urldate = {2025-03-13},
	author = {Maler, E. and Machulak, M. and Richer, J.},
	month = jan,
	year = {2018},
}

@misc{gnap,
	title = {{RFC 9635: Grant Negotiation and Authorization Protocol (GNAP) -- Proposed Standard}},
	url = {https://doi.org/10.17487/RFC9635},
	author = {Richter, J. and Imbault, F.},
	year = {2024},
}

@inproceedings{fedora,
	title = {{Flexible and Extensible Digital Object and Repository Architecture
	         (FEDORA)}},
	author = {Payette, Sandra and Lagoze, Carl},
	editor = {Nikolaou, Christos and Stephanidis, Constantine},
	booktitle = {Research and Advanced Technology for Digital Libraries},
	year = {1998},
	publisher = {Springer Berlin Heidelberg},
	address = {Berlin, Heidelberg},
	pages = {41--59},
	isbn = {978-3-540-49653-3},
}

@misc{lws,
	title = {{Linked Web Storage Working Group Charter}},
	editor = {Pierre-Antoine Champin},
	url = {https://www.w3.org/2024/09/linked-web-storage-wg-charter.html},
	publisher = {World Wide Web Consortium},
	year = {2024},
	month = sep,
}

@techreport{solidlab:uma,
	author = {Termont, Wouter and Dedecker, Ruben and Slabbinck, Wouter and
	          Esteves, Beatriz and De Meester, Ben and Verborgh, Ruben},
	title = {{From Resource Control to Digital Trust with User-Managed Access}},
	year = {2024},
	url = {
	          https://solidlab.be/wp-content/uploads/2024/11/User-Managed-Access-Whitepaper.pdf
	          },
	institution = {{SolidLab} ({IDLab}, {Ghent University} -- {imec})},
}

@inproceedings{dedecker_whats_in_a_pod_2022,
	author = {Dedecker, Ruben and Slabbinck, Wout and Wright, Jesse and
	          Hochstenbach, Patrick and Colpaert, Pieter and Verborgh, Ruben},
	booktitle = {{Proceedings of the QuWeDa 2022 : 6th Workshop on Storing,
	             Querying and Benchmarking Knowledge Graphs co-located with 21st
	             International Semantic Web Conference (ISWC 2022)}},
	issn = {{1613-0073}},
	language = {{eng}},
	location = {{Hangzhou, China (online)}},
	pages = {{81--96}},
	publisher = {{CEUR}},
	title = {{What's in a Pod? A knowledge graph interpretation for the Solid
	         ecosystem}},
	volume = {{3279}},
	year = {{2022}},
	url = {https://ceur-ws.org/Vol-3279/paper6.pdf}
}

@inproceedings{slabbinck_enforcing_usage_control_2024,
	title = {Enforcing {Usage} {Control} {Policies} in {Solid} using {Rule}-{Based} {Web} {Agents}},
	url = {https://ceur-ws.org/Vol-3947/short15.pdf},
	language = {en},
	booktitle = {Proceedings of the {Posters} and {Privacy} {Session} of the {Solid} {Symposium} 2024},
	author = {Slabbinck, Wout and Rojas, Julián Andrés and Esteves, Beatriz and Verborgh, Ruben and Colpaert, Pieter},
	year = {2024},
	pages = {109--117}
}

@electronic{berners_lee_information,
	author = {Berners-Lee, Tim},
	title = {Information management: A proposal },
	url = {http://www.w3.org/History/1989/proposal.html},
	year = 1989,
}

@inproceedings{pandit_dpv,
	author = {J. Pandit, Harshvardhan and Esteves, Beatriz and P. Krog, Georg and
	          Ryan, Paul and Golpayegani, Delaram and Flake, Julian},
	title = {{Data Privacy Vocabulary (DPV) -- Version 2.0}},
	year = {2024},
	isbn = {978-3-031-77846-9},
	publisher = {Springer-Verlag},
	address = {Berlin, Heidelberg},
	doi = {10.1007/978-3-031-77847-6_10},
	booktitle = {The Semantic Web -- ISWC 2024: 23rd International Semantic Web
	             Conference, Baltimore, MD, USA, November 11--15, 2024, Proceedings,
	             Part III},
	pages = {171--193},
	numpages = {23},
	location = {Hanover, MD, USA},
}

@misc{w3c_sai,
	title = {{Solid Application Interoperability: Draft Community Group Report}},
	author = {Bingham, Justin and Prud'hommeaux, Eric and elf Pavlik},
	institution = {World Wide Web Consortium},
	publisher = {Solid Community Group},
	year = {2025},
	url = {https://solidproject.org/TR/sai},
}

@inproceedings{esteves_odrl_profile,
	author = {Esteves, Beatriz and Pandit, Harshvardhan J. and Rodríguez-Doncel,
	          Víctor},
	booktitle = {2021 IEEE European Symposium on Security and Privacy Workshops
	             (EuroS\&PW)},
	title = {{ODRL Profile for Expressing Consent through Granular Access Control
	         Policies in Solid}},
	year = {2021},
	volume = {},
	number = {},
	pages = {298-306},
	doi = {10.1109/EuroSPW54576.2021.00038},
}

@article{taelman_swj_2023,
  title = {Components.js: Semantic Dependency Injection},
  author = {Taelman, Ruben and Van Herwegen, Joachim and Vander Sande, Miel and Verborgh, Ruben},
  journal = {Semantic Web Journal},
  year = 2023,
  volume = 14,
  issue = 1,
  pages = {135--153},
  url = {https://content.iospress.com/articles/semantic-web/sw222945},
  doi = {10.3233/SW-222945},
}

@article{florea_is_2023,
	title = {Is {Automated} {Consent} in {Solid} {GDPR}-{Compliant}? {An} {Approach} for {Obtaining} {Valid} {Consent} with the {Solid} {Protocol}},
	volume = {14},
	issn = {2078-2489},
	doi = {10.3390/info14120631},
	number = {12},
	urldate = {2023-11-30},
	journal = {Information},
	author = {Florea, Marcu and Esteves, Beatriz},
	year = {2023},
}

@phdthesis{esteves_semantic_2024,
	type = {{Doctoral Dissertation}},
	title = {Semantic {Representation} of {Privacy} {Terms} and {Policy}-based {Algorithms} for {Decentralised} {Data} {Environments}},
	copyright = {https://creativecommons.org/licenses/by-nc-nd/3.0/es/},
	url = {https://oa.upm.es/83215/},
	urldate = {2025-07-15},
	school = {E.T.S. de Ingenieros Informáticos (UPM)},
	author = {Esteves, Beatriz},
	collaborator = {Rodríguez Doncel, Víctor and Lewis, David},
	year = {2024},
}

@misc{speicherLinkedDataPlatform2015,
  title = {Linked {{Data Platform}} 1.0},
  author = {Speicher, Steve and Arwe, John and Malhotra, Ashok},
  year = 2015,
  month = feb,
  urldate = {2021-02-23},
  url = {https://www.w3.org/TR/ldp/}
}

@online{capadisli_SolidProtocol_2024,
  title = {{{Solid Protocol}} 0.11.0},
  author = {Capadisli, Sarven and Berners-Lee, Tim and Kjernsmo, Kjetil and Verborgh, Ruben and Bingham, Justin and Zagidulin, Dmitri},
  url = {https://solidproject.org/TR/protocol},
  urldate = {2021-03-29},
  year = 2024
}

@inproceedings{estevesCapturingRequestsContext2025,
	title = {Capturing {Requests} and {Context} for {ODRL}-based {Access} and {Usage} {Control}},
	url = {https://ceur-ws.org/Vol-4093/paper5.pdf},
	booktitle = {Joint {Proceedings} of the 16th {Workshop} on {Ontology} {Design} and {Patterns} and the 1st {Workshop} on {Bridging} {Hybrid} {Intelligence} and the {Semantic} {Web} ({WOP}-{HAIBRIDGE} 2025) co-located with the 24th {International} {Semantic} {Web} {Conference} ({ISWC} 2025)},
	author = {Esteves, Beatriz and Slabbinck, Wout and Sellami, Yassir and Cimmino, Andrea and Rodríguez-Doncel, Víctor and Verborgh, Ruben},
	year = {2025},
}

@inproceedings{bonattiFormalSemanticsOpen2025,
	title = {Towards a {Formal} {Semantics} of the {Open} {Digital} {Rights} {Language} ({ODRL} 2.2)},
	url = {https://ceur-ws.org/Vol-3977/OPAL2025-4.pdf},
	booktitle = {{ODRL} and {Beyond}: {Practical} {Applications} and {Challenges} for {Policy}-based {Access} and {Usage} {Control} ({OPAL} 2025), co-located with the {Extended} {Semantic} {Web} {Conference} 2025 ({ESWC} 2025)},
	author = {Bonatti, Piero Andrea and Fornara, Nicoletta and Harth, Andreas},
	year = {2025},
}

@misc{gdpr,
	title = {Regulation ({EU}) 2016/679 of the {European} {Parliament} and of the {Council} of 27 {April} 2016 on the protection of natural persons with regard to the processing of personal data and on the free movement of such data, and repealing {Directive} 95/46/{EC} ({General} {Data} {Protection} {Regulation})},
	url = {ttp://data.europa.eu/eli/reg/2016/679/oj},
	year = {2016},
}

@misc{data_act,
	title = {Regulation ({EU}) 2023/2854 of the {European} {Parliament} and of the {Council} of 13 {December} 2023 on harmonised rules on fair access to and use of data and amending {Regulation} ({EU}) 2017/2394 and {Directive} ({EU}) 2020/1828 ({Data} {Act})},
	url = {http://data.europa.eu/eli/reg/2023/2854/oj},
	year = {2023},
}
\end{document}